\documentstyle[12pt]{article}

\newcommand{\ignore}[1]{}

\begin{document}
\sloppy
\sloppy
\sloppy
$\ $
\vskip 0.5 truecm
{\baselineskip=14pt
 \rightline{
 \vbox{
       \hbox{UT-772}
       \hbox{April  1997}
}}}
~~\vskip 5mm
\begin{center}
{\large{\bf  Oscillator Quantum Algebra   
                         and    Deformed $su(2)$ Algebra    }}
\end{center}

\vskip .2 truecm
\centerline{\bf Harunobu Kubo\footnote[1]{JSPS fellow,  \qquad 
 E-mail address : kubo@hep-th.phys.s.u-tokyo.ac.jp  } }
\vskip .2 truecm
\centerline {\it Department of Physics,University of Tokyo}
\centerline {\it Bunkyo-ku,Tokyo 113,Japan}

\vskip 0.5 truecm
\begin{abstract}
A difference operator realization of quantum deformed oscillator algebra 
 ${\cal H}_q(1)$ 
with  a Casimir operator freedom is introduced.
We show that this  ${\cal H}_q(1)$  have a nonlinear mapping 
to the deformed quantum $su(2)$ which was introduced by Fujikawa et al. 
We also examine  the  cyclic representation obtained by this 
difference operator realization and the possibility to analyze  
a Bloch electron problem  by    ${\cal H}_q(1)$.
\end{abstract}

Quantum deformed algebra\cite{SDJ} was firstly introduced 
 to study the inverse scattering problems and 
 the integral systems, which   have rich structures, 
Yang-Baxter equations\cite{FKS}.
They are going to be standard techniques of theoretical physics.
Wiegmann and Zabrodin  found that 
a   system of the  Bloch electron  on a two dimensional square lattice
\cite{WZ}\cite{HKW}\cite{FaKa}  
can be expressed in a linear combination of 
generators of the algebra  $U_q(sl_2)$. 
This  corresponds to the fact that 
 $U_q(sl_2)$ gives a foundation to obtain an exact solution of 
the Bethe Ansatz equation.

Bidenharn\cite{B} and Macfarlane\cite{M}
introduced q-deformed oscillator algebra  ${\cal H}_q(1)$  to construct
$U_q(sl_2)$ in the manner of Schwinger's construction of conventional
$su(2)$. 
Recently a new method to construct a representation of   ${\cal H}_q(1)$,  
which is manifestly free of negative norm, was proposed\cite{FKwO}.
This  ${\cal H}_q(1)$ enjoys Hopf structure\cite{OS1}.
This algebra was used to analyze the phase operator problem
\cite{CGN}\cite{E} of the photon with the notion of index \cite{F}. 
Fujikawa et al\cite{FKO}  constructed a new deformation of $su(2)$ algebra 
with a ``Schwinger term'' 
from this non-negative norm representation of  ${\cal H}_q(1)$ 
in the manner of Schwinger's construction of conventional $su(2)$.
We use the ``Schwinger term'' to represent an extra term which deforms the
q-deformed su(2), though  a  better terminology for it may exist. 
It was  confirmed that their algebra coincide with  $U_q(sl_2)$ algebra   
by choosing    specific values of the deformation parameter.

In this paper we introduce a difference operator 
realization of  ${\cal H}_q(1)$ 
with an additional term to prove nonlinear mapping from  
 ${\cal H}_q(1)$  to deformed $su(2)$ with a Schwinger term.
We also study a relation of a Bloch electron problem 
with the difference operator realization of 
deformed oscillator algebra. We show that  
Hamiltonian can be expressed in terms of generators of 
 q-deformed oscillator algebra  and that the Hamiltonian acts on  
 the functional space on which q-deformed oscillator algebra is realized.

We analyze  q-oscillator algebra introduced by Hong Yan\cite{Y}. 
\begin{eqnarray}\label{su2}
[a, a^{\dagger}] &=& [N +1]_q - [N]_q\nonumber\\
{[}N, a^{\dagger}{]} &=& a^{\dagger}\nonumber\\
{[}N, a {]} &=& -a\nonumber\\
{\cal C}_2 &=& a^{\dagger}a - [N]_q
\end{eqnarray}
The usual notation of $[X]_q =\frac{q^{X}-q^{-X}}{q-q^{-1}}$
for deformation parameter $q=\exp(2\pi i \theta)$ with $-1/2< \theta < 1/2$
is used.
Following difference operators,   which  act on the functional space 
$f(x)$,  are shown to realize  this q-oscillator algebra ${\cal H}_q(1)$ .
\begin{eqnarray}\label{Hq}
&&a^{\dagger}f(x)= \mu xf(x) ,\nonumber \\
&&  af(x) = 
 \mu^{-1}\frac{q^{\lambda}f(qx)-q^{-\lambda}f(q^{-1}x)}{(q-q^{-1})x}
    + \mu^{-1}\frac{c} {x}f(x),
 \nonumber \\
&& q^{N}f(x)= q^{\lambda}f(qx), \\
&& q^{-N}f(x)= q^{-\lambda}f(q^{-1}x), \nonumber
\end{eqnarray}
where  $\lambda $,  $\mu$ and  $c$ are undetermined parameters.
In the case special of  $\lambda =0$,  $\mu =1$ and  $c=0$,  
this difference realization coincide that of Ref.\cite{Y}. 
In the  realization (\ref{Hq}) the Casimir operator ${\cal C}_2$ has a value,
\begin{equation}
{\cal C}_2f(x)=
     ( a^{\dagger}a - [N])f(x)=( a a^{\dagger} - [N+1])f(x)= cf(x).
\end{equation}
We note that  when $q$ is a root of unity we have additional central elements  
of  ${\cal H}_q(1)$  and their values 
depend on  parameters  $\lambda$  and $\mu $. Then we can understand that
the defining relations (\ref{Hq}) of  ${\cal H}_q(1)$ on the functional space contain
the degrees of freedoms which will correspond to the values 
of the central elements.

On this functional space we can construct  a 
Fock space with generic parameters  
$\lambda$, $\mu $ and $c$.
The operator $N$ and $a^{\dagger}a $ commute with each other, 
and thus they  can be 
diagonalized simultaneously  on the state $|\psi_0\rangle$\cite{OS2},
\begin{equation}\label{ground}
N|\psi_0\rangle = \nu_0|\psi_0\rangle, \qquad
a^{\dagger}a |\psi_0\rangle = \lambda_0|\psi_0\rangle. 
\end{equation}
Our representation of  ${\cal H}_q(1) $ 
has  a Casimir operator 
${\cal C}_2$, and  
we obtain an  additional constraint 
$c=\lambda_0 - [\nu_0]_q$ by acting ${\cal C}_2$ on $|\psi_0\rangle$.
We  impose  a  highest weight state condition $  \lambda_0 = 0$,
and  we can determine  $\nu_0$ by $c$,
\begin{equation}\label{hws}
c= - [\nu_0]_q .
\end{equation}
All the states are  generated by 
applying $a^{\dagger}$ on the highest weight state,
$ |\psi_n \rangle =a^{\dagger n}|\psi_{0}\rangle$.
We have for $n \ge 0$, 
\begin{equation}\label{lambda}
N|\psi_{n} \rangle =(\nu_0 + n)|\psi_{n}\rangle, \quad  
  a^{\dagger}|\psi_n \rangle =|\psi_{n+1}\rangle, 
  \quad
  a |\psi_{n}\rangle =\lambda_{n} |\psi_{n-1}\rangle, 
  \quad \lambda_{n} =  [n+\nu_0]_q - [\nu_0]_q
\end{equation}
Those Fock states  $|\psi_n \rangle $ can be constructed explicitly 
on the functional space up to normalization,
\begin{equation}\label{func}
|\psi_n \rangle \sim \mu^{n} x^{\nu_0 - \lambda +n }. 
\end{equation}
In fact, the  equations (\ref{ground}) and (\ref{lambda}) are easily checked by using 
$N = x\partial_x + \lambda$, 
which is derived from difference operator realization in  (\ref{Hq}).
This Fock space  certainly  contains  an  ordinary  representation of 
 ${\cal H}_q(1) $. But if we want to construct a cyclic representation
 from a finite dimensional irreducible representation of ${\cal H}_q(1) $,
 which can be obtained for   $q$ being a root of unity, we ought to clear
some subtle points. This is because in order to obtain a cyclic 
representation,  the Fock space should have integral powers in $x$.
We thus need to examine the functional space (\ref{func}), whose bases do  
 not always have integral powers in $x$. 

We can have a finite dimensional irreducible  Fock space,
if a state is truncated as 
$aa^{\dagger}| \psi_k \rangle =a| \psi_{k+1} \rangle 
=\lambda_{k+1}| \psi_k \rangle = 0$.
So, it is enough to examine a condition $\lambda_{k+1} =0$.
For the generic  $ \theta$,  we can have a finite dimensional irreducible
representation by choosing a suitable value of $\nu_0$ \cite{OS2}.
Let us consider 
the rational $\theta = P/Q$, where $P$ and $Q$ are relatively prime.
The condition  $\lambda_{k+1} =0$,  which determines 
if the truncation of states occur or not, is
\begin{equation}\label{cos}
\cos\left(\pi \frac{P}{2Q}(2\nu_0 + k + 1)\right) 
\sin\left(\pi \frac{P}{2Q}(k+1)\right) = 0 . 
\end{equation}
Then we can have a finite irreducible representation if this condition is 
satisfied. 
The cyclic representation of the  algebra (\ref{Hq}) can be obtained with the 
condition (\ref{cos}), by  setting 
$x=q^{k}$ in (\ref{Hq}).  The  cyclic basis are defined by,
\begin{equation}\label{cycB}
f_{k}= f(q^{k}), \qquad   f_{k + 2Q}=f_k.
\end{equation}
Only the functions which satisfy $f(q^{2Q}x)=f(x)$ are admissible to construct
this cyclic basis. This means that the function $f(x)$  should consist of  
integral powers of $x$,  that is,
\begin{equation}\label{Z}
\nu_0 - \lambda \in  Z, 
\end{equation}
in (\ref{func}).        
If the above conditions (\ref{cos}) and (\ref{Z}) 
are satisfied we can have 
a cyclic  representation $\rho$ of  ${\cal H}_q(1)$ 
by inserting  $x =q^{k}$ in (\ref{Hq}).
\begin{eqnarray}\label{osc-cyclic}
&&\rho(a^{\dagger})f_k= \mu q^{k}f_k , \nonumber \\
&& \rho( a)f_k =  \mu^{-1} (q-q^{-1})^{-1}q^{-k}
\left(q^{\lambda} f_{k+1}-q^{-\lambda} f_{k-1}\right)
+ \mu^{-1}c q^{-k}f_k, \nonumber \\
&& \rho(q^{N})f_k= q^{\lambda}f_{k+1} , \\
&& \rho(q^{-N})f_k= q^{-\lambda}f_{k-1}. \nonumber
\end{eqnarray}
Let   ${\cal A}$ be  a cyclic representation of the algebra, then 
 ${\cal A}$ has
following properties.   
\begin{equation}
\rho( \alpha a + \beta b)\vec{f}= \alpha \rho(a)\vec{f} + 
 \beta \rho(b)\vec{f}, \qquad
\rho(  a b)\vec{f}=  \rho(b)  \rho(a) \vec{f},  
\end{equation}
where  $a,b \in {\cal A}$,   $\alpha , \beta \in  C$ and    cyclic bases 
$\vec{f}=(f_1,\cdots , f_{2Q})^{t}$.
This cyclic representation of  the generators of ${\cal H}_q(1)$ 
can be expressed by using Weyl bases $X$ and $Y$.
\begin{eqnarray}\label{HqW}
&&\rho(a^{\dagger})\vec{f}= \mu Y\vec{f},  \nonumber \\
&& \rho(a)\vec{f}=\mu^{-1}
  (q-q^{-1})^{-1}Y^{-1}\left(q^{\lambda} X^{-1}-q^{-\lambda} X\right)\vec{f} 
  + \mu^{-1}cY^{-1}\vec{f}, \nonumber \\
&& \rho(q^{N})\vec{f}= q^{\lambda}X^{-1}\vec{f},\\
&& \rho(q^{-N})\vec{f}= q^{-\lambda}X\vec{f}  . \nonumber 
\end{eqnarray}
We have used the following  $2Q \times 2Q$ matrix realization\cite{HKW}
of  Heisenberg-Weyl group,
\begin{equation}\label{example}
X=\left[
\begin{array}{cccccc}
0 &       &      &      &      & 1 \\
1 &   0   &      &    &      &   \\
  &   1   &  \cdots &    &      &   \\
  &       &        & \cdots   & 0    &   \\
  &       &        &        & 1   &  0 
\end{array}
\right],
\quad Y= \mbox{diag}(q, \cdots , q^{2Q})
\end{equation}
In this realization $X$ and $Y$ satisfy  $qXY = YX$ 
and  $Y^{2Q} = X^{2Q} =1 $.

Next we will show that 
 the difference operator realization (\ref{Hq}) of ${\cal H}_q(1)$ can
be mapped to a deformed $su(2)$ algebra  with a Schwinger term.
This  deformed $su(2)$ algebra is  defined by the following relations\cite{FKO},
\begin{eqnarray}\label{defSn}
{[}S_{3}, S_{\pm} {]} &=& \pm S_{\pm},\nonumber\\
{[}S_{+}, S_{-}{]}&=& [2S_{3}]_q + \xi [S_{3}]_q.
\end{eqnarray}
The last term in (\ref{defSn}) which is proportional to the $\xi$, 
gives rise to an extra term in the conventional 
q-deformed $su(2)$ algebra. 
This extra term play an essential role to construct a  positive norm 
representation,  and it has been tentatively called ``Schwinger term''.
If $\xi=0$  holds, this algebra is the  
same as the conventional q-deformed $su(2)$.
The $2j+1$ dimensional highest weight representation of 
the algebra (\ref{defSn}) can also be realized by q-difference equations as 
\begin{eqnarray}\label{defS}
S_{+}f(x)&=&(q\!-\!q^{-1})^{-1}x(q^{2j-n_{0}-\kappa}f(q^{-1}x)
\!-\!q^{-2j+n_{0}+\kappa}f(qx))  +x[n_0]f(x), \nonumber \\
S_{-}f(x)&=&-(q\!-\!q^{-1})^{-1}x^{-1}(q^{n_{0}-\kappa}f(q^{-1}x)
\!-\!q^{-n_{0}+\kappa}f(qx))  +x^{-1}[n_0]f(x),  \\
q^{S_{3}}f(x)&=&q^{\kappa - j}f(qx), \nonumber  \\
\xi &=& (q^{\frac{1}{2}}-q^{-\frac{1}{2}})(q-q^{-1})
[n_{0}]_q [n_{0}-j-\frac{1}{2}]_q,   \nonumber
\end{eqnarray}
where we introduced  $n_0$ to parameterize  $\xi$, and 
$\kappa$ is an  undetermined parameter. 
This representation satisfies 
the highest weight condition $\tilde{S}_{+}x^{2j-\kappa}=0$ 
and the lowest weight condition $\tilde{S}_{-}\cdot x^{-\kappa} =0$. 
We have $\{ x^{-\kappa},  x^{-\kappa +1}, \cdots , x^{-\kappa +2j}\}$
for bases of this $2j+1$ dimensional representation.

The deformed $su(2)$ with a Schwinger term (\ref{defSn}) is 
{\it non-linearly} realized  by  using the generators of 
q-deformed oscillator ${\cal H}_q(1)$.
Comparing difference operators in (\ref{Hq}) and (\ref{defS}), 
we can  understand that there exist following 
realizations  of $S_+$,  $S_-$ and $S_3$  
 in terms of the generators of  ${\cal H}_q(1)$,
\begin{eqnarray}\label{ans}
&& S_+=(-[N+\delta]_q +c)a^{\dagger},   \nonumber \\
&& S_-=a,  \\
&& S_3=N + \gamma ,  \nonumber
\end{eqnarray}
where the  overall factor $\mu$ in (\ref{Hq}) 
can always be removed in defining relations
 (\ref{defS}) of deformed $su(2)$, and thus we can set $\mu=1$.   
We also make the   following identifications,
\begin{equation}\label{choice}
   c = [n_0]_q,  \qquad 
   q^{\lambda}=q^{-n_0 + \kappa},   
   \qquad q^{\delta}=q^{2n_0 - 2j -1 +\kappa},
   \qquad q^{\gamma }=q^{n_0-j +\kappa}.
\end{equation}
Inserting  this choice (\ref{choice}) of parameters  into (\ref{ans})
we have  the difference operator realizations of q-deformed $su(2)$.  
Parameters $c$ and $\lambda$ of ${\cal H}_q(1)$, and also   
$\delta$ and $\gamma$ introduced to construct the mapping 
are determined by the data of 
the $2j+1$ dimensional representation of deformed $su(2)$, namely 
$n_0$, $j$ and  $k$.
This mapping (\ref{choice}) is  different from 
the original realization 
of the deformed $su(2)$ which was  constructed from  two kinds of  q-deformed 
oscillators\cite{FKO}.

We next  study if the difference operator realization of  ${\cal H}_q(1)$ can 
be used to analyze a Bloch electron on the square lattice.
A Hamiltonian with two anisotropic parameters $V_1$ and $V_2$ 
can be written on the cyclic basis (\ref{cycB}) with momentum $p_{\pm}$ 
by using  $X$ and $Y$ \cite{HKW}, 
\begin{equation} \label{HamiltonianXYv}
H=  V_{1}e^{i(p_+ + p_-)}Y^{-1}X^{-1} 
+ V_{2} e^{i(p_+ - p_-)}X^{-1}Y
+  V_{1}e^{-i(p_+ + p_-)}XY
+ V_{2} e^{-i(p_+ - p_-)}Y^{-1}X.
\end{equation}
We examine  if this Hamiltonian  can be expressed in terms of the 
generators of ${\cal H}_q(1)$ which realize the cyclic representation  
(\ref{osc-cyclic}).
We can use  a  following  ansatz,
\begin{eqnarray}\label{HamiltonianXYe}
H&=& \epsilon_1 \rho(a) + \epsilon_2 \rho(q^{N})\rho( a^{\dagger})
 +\epsilon_3  \rho(q^{-N})\rho( a^{\dagger}) 
 \\
&=& \epsilon_1 \mu^{-1}
(q-q^{-1})^{-1}Y^{-1}\left(q^{\lambda} X^{-1}-q^{-\lambda} X\right)
 + \epsilon_1 \mu^{-1}cY^{-1} 
+\epsilon_2 q^{\lambda}X^{-1} \mu Y +\epsilon_3  q^{-\lambda}X  \mu Y.
\nonumber
\end{eqnarray}
Comparing those two expressions of Hamiltonian (\ref{HamiltonianXYv}) and 
(\ref{HamiltonianXYe}),   we have conditions for $ \epsilon_1$,
 $ \epsilon_2$, $ \epsilon_3$ and $c$, 
\begin{eqnarray}\label{c}
&& \epsilon_1 \mu^{-1}(q-q^{-1})^{-1}q^{\lambda}= V_{1}e^{i(p_+ + p_-)},
\quad
-\epsilon_1 \mu^{-1}(q-q^{-1})^{-1}q^{-\lambda}=V_{2} e^{-i(p_+ - p_-)},
\nonumber \\
&&\epsilon_2 q^{\lambda}\mu = V_{2} e^{i(p_+ - p_-)},
\quad 
\epsilon_3 q^{-\lambda} \mu = V_{1}e^{-i(p_+ + p_-)},
\quad
 c = 0.
\end{eqnarray}
In order to  keep the generators of  ${\cal H}_q(1)$ 
certainly corresponding to the cyclic representation,
we also need the condition (\ref{Z}).
We should  check if  $c=0$ (\ref{c}) and   $\nu_0 -\lambda \in Z$ (\ref{Z})   
conditions are satisfied or not. 
$c=0$ means $q^{2\nu_0}=1$ from (\ref{hws}), 
and   $\nu_0 -\lambda \in Z$ means 
$q^{2\lambda}=q^{2\nu_0  +2\kappa}=q^{2\kappa}$ with $\kappa \in Z$.
Then we have the condition,
\begin{equation}\label{midband}
\left( -\frac{V_1}{V_2}\right) e^{2ip_+}=q^{2\kappa}.
\end{equation}
This condition (\ref{midband}) leads to an isotropic and a midband condition,
$V_1 = V_2$, and $p_+ = \pi/2$ mod $\pi/Q$. 
The first two  equations in (\ref{c})  
can be satisfied simultaneously by
$- e^{2 ip_+} = q^{2\lambda}$.  
Then we can determine the parameter $\lambda$ as 
$q^{\lambda}= \pm i e^{ ip_+}$.
Let us insert this solutions for  $\epsilon_1,\epsilon_2$ and $ \epsilon_3$ 
into the  Hamiltonian  (\ref{HamiltonianXYe}),
\begin{equation}\label{Ha}
H =   i (q-q^{-1})\left(
      \rho(a)+\rho([N ]_q)\rho( a^{\dagger})
      \right).
\end{equation}
where we  set    $e^{ ip_-}\mu=1$ and  $V_1 =1$ for simplicity 
and choose plus sign in (\ref{Ha}) 
which comes from the  condition $q^{\lambda}= \pm i e^{ ip_+}$.
By considering  eq (\ref{cos}), 
we can find that 
the Fock space  is truncated at $k=Q-1$ for $P=$ odd and even, 
and  the Fock space becomes  the $Q$ dimensional representation space 
 $|\psi_i \rangle, i=0,1,\cdots k=Q-1$.
This means that Hamiltonian which is realized by ${\cal H}_q(1)$ on the
functional space with  integral powers 
certainly corresponds to a  cyclic representation 
by inserting $x=q^k$ as  usual way.

Because we constructed the relations (\ref{ans})
between   ${\cal H}_q(1)$  and deformed $su(2)$
with a Schwinger term, the Hamiltonian (\ref{Ha})  can be  rewritten 
in terms of  deformed $su(2)$, $H=-(q-q^{-1})(S_- - S_+)$ with a suitable 
choice of $\kappa$, say $\kappa = 0$. 
We have constraints $[n_0]=c=0$ by (\ref{choice}) and (\ref{c}), 
then the Schwinger term in deformed $su(2)$ vanishes. 
Then $S_+$, $S_-$ and $S_3$ form  the  conventional q-deformed $su(2)$.

We note that since the deformation parameter $q=\exp(i\pi P/Q)$ satisfies 
$q^{2Q}=1$, we can rewrite the Hamiltonian (\ref{Ha}) in another form,
\begin{equation}\label{another}
H =i(q-q^{-1})\left(
     \rho( a)+q^{Q}\rho([N +Q]_q )\rho(a^{\dagger})
      \right).
\end{equation}
Using this alternative  expression of the Hamiltonian (\ref{another}),
 we can introduce  generators of  $U_q(sl_2)$, by  
$B = -[N+Q]_q a^{\dagger}, C = a, A = q^{N + \frac{1+Q}{2}}$ 
and $ D=q^{-N - \frac{1+Q}{2}}$. 
The Hamiltonian can be written  in this identification
$H=i(q-q^{-1})\left( C \pm  B   \right),  \quad P=\mbox{odd, even}$,
respectively.
This expression of Hamiltonian  has  been  found  by  Wiegmann and Zabrodin to 
obtain the exact results of Bethe Ansats equation\cite{WZ}\cite{HKW}. 

In conclusion, 
we introduced a difference operator realization of
  ${\cal H}_q(1)$ with parameters
which correspond to the degrees of freedom of central elements 
in the case of $q$ being 
a root of unity and the value of the Casimir ${\cal C}_2$.
 This realization of ${\cal H}_q(1)$ gives {\it non-linear} mapping
form  ${\cal H}_q(1)$ to 
the deformed $su(2)$ with a Schwinger term 
which was introduced by Fujikawa et al\cite{FKO}.
In the 
case of Schwinger term  vanishing, we can deduce a correspondence between 
 ${\cal H}_q(1)$ and the conventional q-deformed $su(2)$.
We also examined if this difference realization of  ${\cal H}_q(1)$ can 
describe the Hamiltonian of a Bloch electron problem
 on the square lattice.
We found that 
the Hamiltonian has an  expression  in terms of generators of  ${\cal H}_q(1)$
acting on the functional space.
Using  non-linear mapping form  ${\cal H}_q(1)$ to the deformed $su(2)$, 
we can find  that the Hamiltonian  (\ref{another})
coincide with what was found by  Wiegmann and Zabrodin.
We hope that the q-deformed oscillator algebra which 
realizes the Hamiltonian (\ref{Ha}) gives further 
information about  the  Bloch electron problem in the future, 
since new mathematical  machinery  such as the   coherent   representation,
etc., may become available.    
   
{\bf Acknowledgments } We would like to thank 
K. Fujikawa, Y. Hatsugai, Y. Morita,  
J. Shiraishi  for  valuable discussions. 


\begin{thebibliography}{99}
\bibitem{SDJ}
E. K. Sklyanin, Usp. Mat. Nauk {\bf 40}, 214(1985).\\
V. G. Drinfeld, Dokl. Acad. Nauk {\bf 283}, 1060(1985).\\
M. Jimbo, Lett. Math. Phys. {\bf 10}, 63(1985).
\bibitem{FKS}
L. D. Faddeev, {\em Les Houches Lectures} 1982(Elsevier, Amsterdam, 1984).\\
P. P. Kulish and E. K. Sklyanin,\  {\em Lecture Note in Physics} Vol. 151(Springer, Berlin,1982).
\bibitem{B}
L. C. Biedenharn,J. Phys. A: Math. Gen. {\bf 22}, L873(1989). 
\bibitem{M}
A. J. Macfarlane, J. Phys. A: Math. Gen. {\bf 22}, 4581(1989).
\bibitem{WZ}
P. B. Wiegmann and A. V. Zabrodin, Phys. Rev. Lett. {\bf 72}, 1890(1994); Nucl.
Phys. {\bf B422}[FS], 495(1994).
\bibitem{HKW}
Y. Hatsugai, M. Kohmoto, and Y. S. Wu, Phys. Rev. Lett. {\bf 73}, 1134(1994);cond-mat/9509062.
\bibitem{FaKa}
L. D. Faddeev and R. M. Kashaev, Comm. Math. Phys. {\bf 169},181(1995).
\bibitem{FKO}
K. Fujikawa, H. Kubo, C. H. Oh,   Mod. Phys. Lett. A {\bf 12}, 403(1997).     
\bibitem{CGN}
S. H. Chiu, R. W. Gray and C. Nelson, Phys. Lett. {\bf A164}, 237(1992).
\bibitem{E}
D. Ellinas, Phys. Rev. {\bf A45}, 3358(1992).
\bibitem{Y}
Hong Yan, J. Phys. A:Math. Gen. {\bf 23}, L1150(1990).
\bibitem{FKwO}
K. Fujikawa, L. C. Kwek, and C. H. Oh, Mod. Phys. Lett. {\bf A10}, 2543(1995).
\bibitem{F}
K. Fujikawa, Phys. Rev. {\bf A52},3299(1995).
\bibitem{OS1}
C. H. Oh and K. Singh, J. Phys A:Math. Gen. {\bf 27}, 5907(1994).
\bibitem{OS2}
C. H. Oh and K. Singh,  Lett. Math. Phys. {\bf 36}, 77(1996).
\end{thebibliography}
\end{document}